\documentclass{article}

\usepackage{PRIMEarxiv}

\usepackage[utf8]{inputenc} 
\usepackage[T1]{fontenc}    
\usepackage{hyperref}       
\usepackage{url}            
\usepackage{booktabs}       
\usepackage{amsfonts}       
\usepackage{nicefrac}       
\usepackage{microtype}      
\usepackage{lipsum}
\usepackage{fancyhdr}       
\usepackage{graphicx}       
\graphicspath{{media/}}     

\title{Trustworthy modelling of atmospheric formaldehyde powered by deep learning
}

\author{
  Mriganka Sekhar Biswas \\
  Indian Institute of Tropical Meteorology \\
  Pune, India\\
  \texttt{mriganka@tropmet.res.in} \\
   \And
  Manmeet Singh \\
  Indian Institute of Tropical Meteorology \\
  Pune, India\\
  \texttt{manmeet.cat@tropmet.res.in} \\
}

\begin{document}
\maketitle

\begin{abstract}
Formaldehyde (HCHO) is one one of the most important trace gas in the atmosphere, as it is a pollutant causing respiratory and other diseases. It is also a precursor of tropospheric ozone which damages crops and deteriorates human health. Study of HCHO chemistry and long-term monitoring using satellite data is important from the perspective of human health, food security and air pollution. Dynamic atmospheric chemistry models struggle to simulate atmospheric formaldehyde and often overestimate by up to two times relative to satellite observations and reanalysis. Spatial distribution of modelled HCHO also fail to match satellite observations. Here, we present deep learning approach using a simple super-resolution based convolutional neural network towards simulating fast and reliable atmospheric HCHO. Our approach is an indirect method of HCHO estimation without the need to chemical equations. We find that deep learning outperforms dynamical model simulations which involves complicated atmospheric chemistry representation. Causality establishing the nonlinear relationships of different variables to target formaldehyde is established in our approach by using a variety of precursors from meteorology and chemical reanalysis to target OMI AURA satellite based HCHO predictions. We choose South Asia for testing our implementation as it doesnt have in situ measurements of formaldehyde and there is a need for improved quality data over the region. Moreover, there are spatial and temporal data gaps in the satellite product which can be removed by trustworthy modelling of atmospheric formaldehyde. This study is a novel attempt using computer vision for trustworthy modelling of formaldehyde from remote sensing can lead to cascading societal benefits.
\end{abstract}


\section{Introduction}
\label{sec:intro}
Trace gases are of utmost importance in atmospheric chemistry as some of them are potential pollutants (e.g. \(NO_2\), \(SO_2\), HCHO etc.) \cite{U.S.EPA2011}. They pose health risk for humans, reduce agricultural yields and affect the complete biosphere. Some other trace gases act as greenhouse gases (GHGs) affecting atmospheric radiation budget and contribute towards climate change \cite{IPCC2022,fadnavis2020atmospheric}. Formaldehyde (HCHO), the most abundant aldehyde in the atmosphere, is also a pollutant causing respiratory diseases and other complications for the health of homo sapiens \cite{U.S.EPA2011}. The most important role of HCHO in atmospheric chemistry is that it acts as a precursor for tropospheric ozone \(O_3\) \cite{Crutzen1974,CarterAtkinson1987}. Although stratospheric \(O_3\) protects all the life-forms on the Earth by absorbing harmful ultraviolet (UV) radiations and preventing the same from entering Earth's atmosphere, the tropospheric \(O_3\) is a pollutant. Tropospheric ozone is currently responsible for a total loss of  4-15\%, 6-16\%, 3-4\%, and 2.2 - 5.5\% for wheat, soybean, rice and maize respectively, with global estimated crop loss of ~\$11 - 26 billion \cite{mills2011ozone}. Short-term \(O_3\) exposure is related to respiratory and cardiovascular disease, eventually leading to premature mortality \cite{world2021review,brown2013integrated}. Hence, the study of atmospheric \(HCHO\) is important from atmospheric chemistry, air pollution, food security and public health point of view.

Oxidation of methane (\(CH_4\)) and higher volatile organic compunds (VOCs) are the main sources of tropospheric HCHO. Precursors of HCHO (\(CH_4\) and higher VOCs) can be biogenic, anthropogeic or pyrogenic \cite{carlier1986chemistry,zhu2017long,de2010trend}. Vegetation, primarily forests emit various VOCs including isoprenes (\(C_5H_4\)) and monoterpenes at higher temperature (~30 degree C) \cite{simpson1995biogenic}. Dynamic atmspheric chemistry models use different algorithms to calculate VOC emission rates with respect to atmospheric temperature \cite{lamb1987national, simpson1995biogenic}. These VOCs get oxodise in presence of solar radiation to form HCHO and in the process, tropospheric \(O_3\) is being generated \cite{biswas2019simultaneous}. The background concentration of atmospheric HCHO is maintained by the oxidation of (\(C_5H_4\)), whereas the other VOCs are responsible for the spatial and temporal variability of atmospheric HCHO. Among the biogenic VOCs, (\(C_5H_4\)) is the largest contributor for HCHO concentration. Various previous studies have reported estimation of \(C_5H_4\) emission from HCHO satellite obervations \cite{millet2006formaldehyde,palmer2003mapping,palmer2006quantifying,wolfe2016formaldehyde}. As HCHO is being formed in the atmosphere from the oxidation of VOCs, it has been used as an indicator for biogenic and anthropogenic VOC emission \cite{andreae2001emission,stavrakou2009evaluating}. 

\subsection{Chemical reactions of formaldehyde and related chemical species}
Reactions of formaldehyde, VOCs and other related chemical species happens in tandem and cannot be discussed separately as these reactions are interdependent. The photochemical reactions are fueled by specific wavelength of solar radiation based on the chemical properties of the reactants. Nitrogen dioxide \(NO_2\) acts as catalyst for HCHO and VOC oxidation leading to more formaldehyde and \(O_3\) and eventually regenerating again. \(NO_2\) first gets photodissociated in presence of solar radiation to form \(NO\) and activated oxygen radical (\(O(^3P)\)).
\begin{equation}
NO2 + hv -> NO + O(^3P)
\end{equation}
\(O^3P\) reacts with oxygen molecule to form ozone while \(NO_2\) is regenerated from \(NO\).
\begin{equation}
O(^3P) + O2 + M -> O3 + M
\end{equation}
\begin{equation}
HO2 + NO -> NO2 + OH
\end{equation}
Ozone also photodissociates to form another form of oxygen radical (\(O(^1D)\)) which reacts with water molecule to form the an important atmospheric oxidant, \(OH\) radical.
\begin{equation}
O3 + hv -> O(^1D) + O2
\end{equation}
\begin{equation}
O(^1D) + H2O -> 2 OH
\end{equation}
Methane and other hydrocarbons get oxidised by \(OH\) radical to form formaldehyde (for simplicity reactions with only methane is presented here).
\begin{equation}
OH + CH4  -> CH3 + H2O
\end{equation}
\begin{equation}
CH3 + O2 + M -> CH3O2 + M
\end{equation}
\begin{equation}
CH3O2 + NO -> CH3O + NO2
\end{equation}
\begin{equation}
CH3O + O2 -> HCHO + HO2
\end{equation}
Finally the formaldehyde photooxidised to carbon monoxide along with the formation of \(HO_2\)
\begin{equation}
HCHO + hv -> H + HCO
\end{equation}
\begin{equation}
HCHO + OH -> H2O + HCO
\end{equation}
\begin{equation}
HCO + O2 -> CO + HO2
\end{equation}

\subsection{Previous studies on atmospheric formaldehyde}
\begin{figure*}[t]
  \centering
   \includegraphics[width=0.8\linewidth]{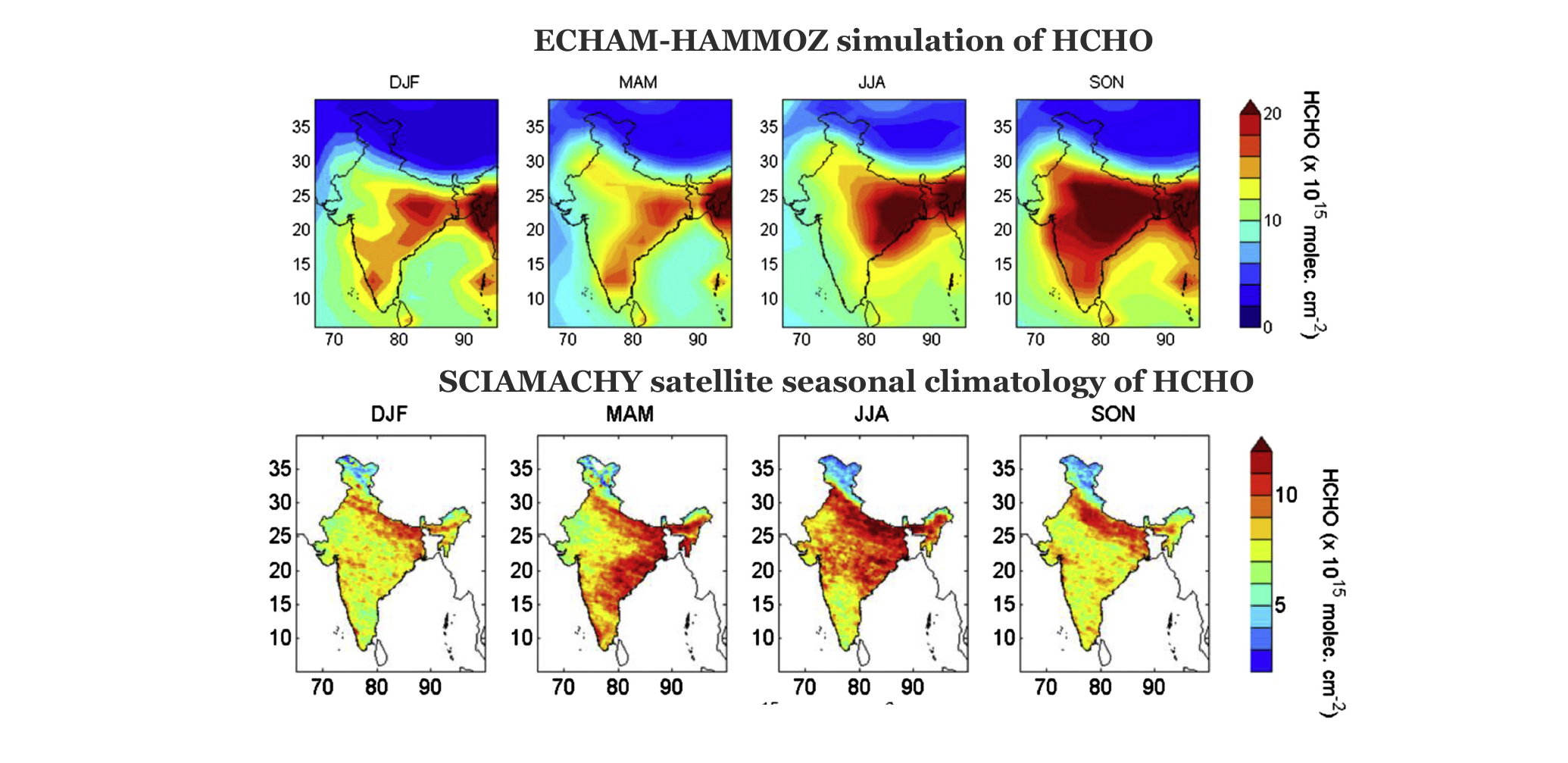}

   \caption{Spatial distributions of formaldehyde (HCHO) from the dynamical model ECHAM-HAMMOZ (top row) and SCIAMACHY satellite (bottom). The figure represents seasonal climatology over South Asia. Difference in absolute value and spatial distribution between modelled and satellite measured formaldehyde is evident. The deviation is most prominent during Sept-Oct-Nov period. (with permission from \cite{mahajan2015inter} ).}
   \label{fig:onecol}
\end{figure*}

Satellite based long-term studies have reported spatial variation and temporal evolution of HCHO. Smedt et. al. have reported a 1.6\% increase per year over India using SCIAMACHY satellite observations \cite{de2010trend}. A later study estimated an increase of 1.51 ± 0.44\% HCHO vertical column density (VCD) per year over India \cite{mahajan2015inter}. Observation based studies have reported that HCHO concentrations over Indian cities are dominated by anthropogenic emissions with characteristic signal from automobile emission \cite{biswas2021year}, whereas over rural and high-alitude regions the HCHO concentrations are dominated by biogenic emission and boundary layer evolutions respectively \cite{biswas2019simultaneous,biswas2021effect}. Using GEOS-Chem model simulation to interpret OMI satellite observations of HCHO, \cite{surl2018processes} have reported that HCHO seasonal cycle over India correlate with surface temperature. They concluded that the biogenic emissions drive HCHO concentrations over India. \cite{chaliyakunnel2019constraining} compared GEOS-Chem simulation with satellite (OMI, GOME-2a) HCHO and reported that biogenic VOC emissions are ~30-60\% overestimated. However, past studies have also reported significant mismatch between dynamic model simulation and satellite observations of HCHO. \cite{mahajan2015inter} report that ECHAM5-HAMMOZ model overestimates HCHO compared to satellite observations. In figure \ref{fig:onecol} we have presented comparison between ECHAM5-HAMMOZ model and SCIAMACHY satellite observations from \cite{mahajan2015inter} (reproduced with permission). The modelled HCHO is much larger (~2 times) as compared to the satellite observations. Moreover, significant differences between spatial distribution and absolute values can be seen. \cite{chutia2019distribution} report up to ~40\% overestimation in modelled (WRF-Chem) HCHO compared to MACC reanalysis. 

\subsection{Machine learning based studies on atmospheric chemistry}

The use of machine learning for atmospheric chemistry is still in a nascent stage. \cite{ojha2021exploring} demonstrate the use of machine learning for low-cost simulations of urban ozone concentration fluctuations in the Himalayan foothills region. They are able to generate reasonable seasonal cycles of O3 and climate characteristics. In the absence of high-resolution measurements, they show that machine learning can provide alternate intelligent estimates of ozone important for the societal well being. \cite{guan2021global} construct confidence intervals for the 2019 global surface HCHO at multiple confidence levels using neural networks. Their data-driven approach, shows that over Southeast Asia, North China, Central and Western Africa, and Latin America's rainforest areas, all had much greater levels of HCHO levels than the rest of the world. With the menace of air pollution increasing on a global basis, the applications of machine learning in atmospheric chemistry have a vast opportunity.

\begin{figure*}[t]
  \centering
   \includegraphics[width=0.8\linewidth]{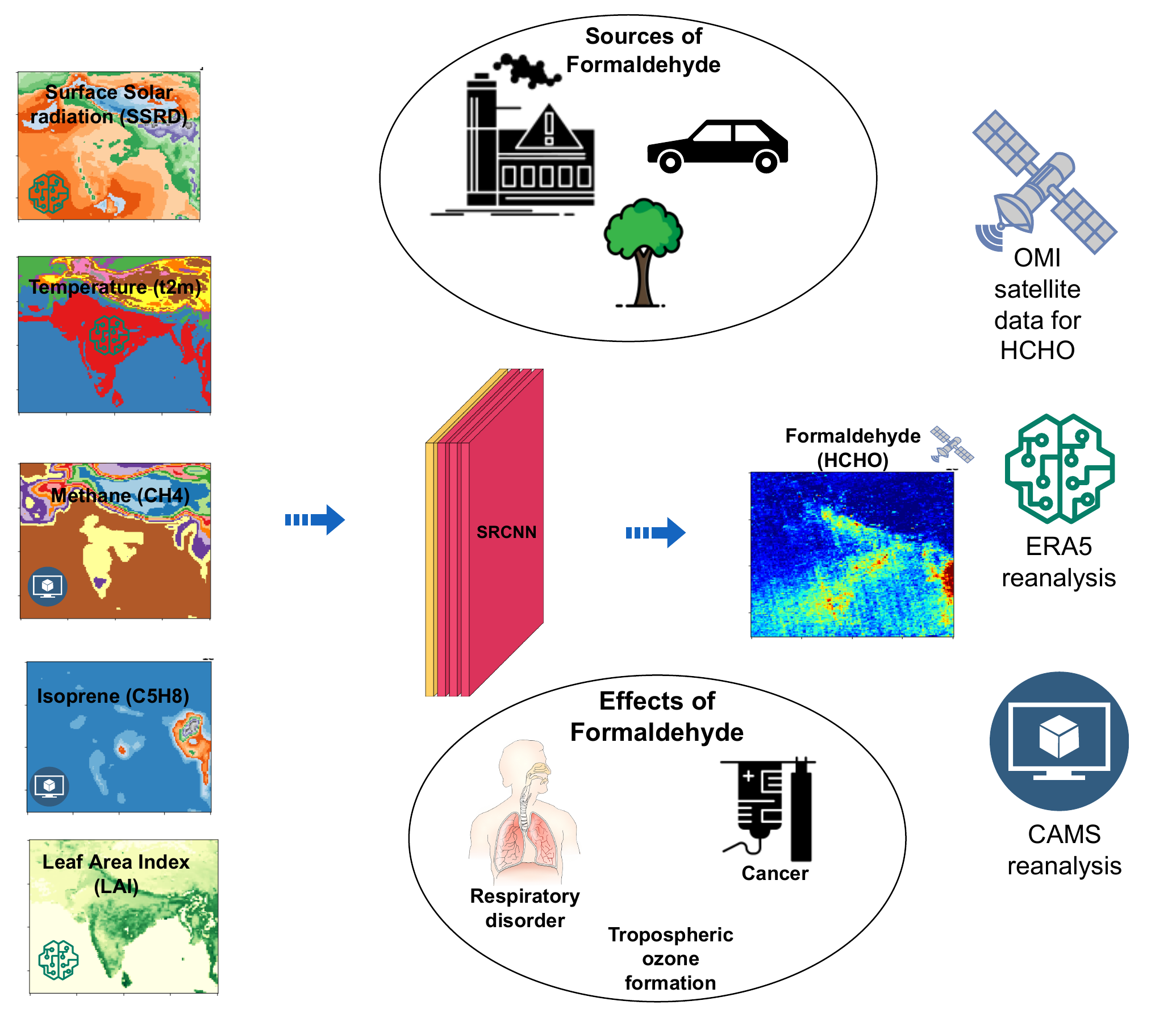}

   \caption{Schematic diagram of the present work. Formaldehyde (HCHO) is a pollutant and responsible for respiratory diseases, cancer and ozone pollution. Biogenic and anthropogeic VOCs oxidise (thermally and photochemically) to form atmospheric HCHO. Using atmospheric temperature, incoming solar radiation, methane, isoperene and higher VOC concentrations as input we have simulated HCHO.}
   \label{fig:two}
\end{figure*}
\begin{figure*}[t]
  \centering
   \includegraphics[width=0.8\linewidth]{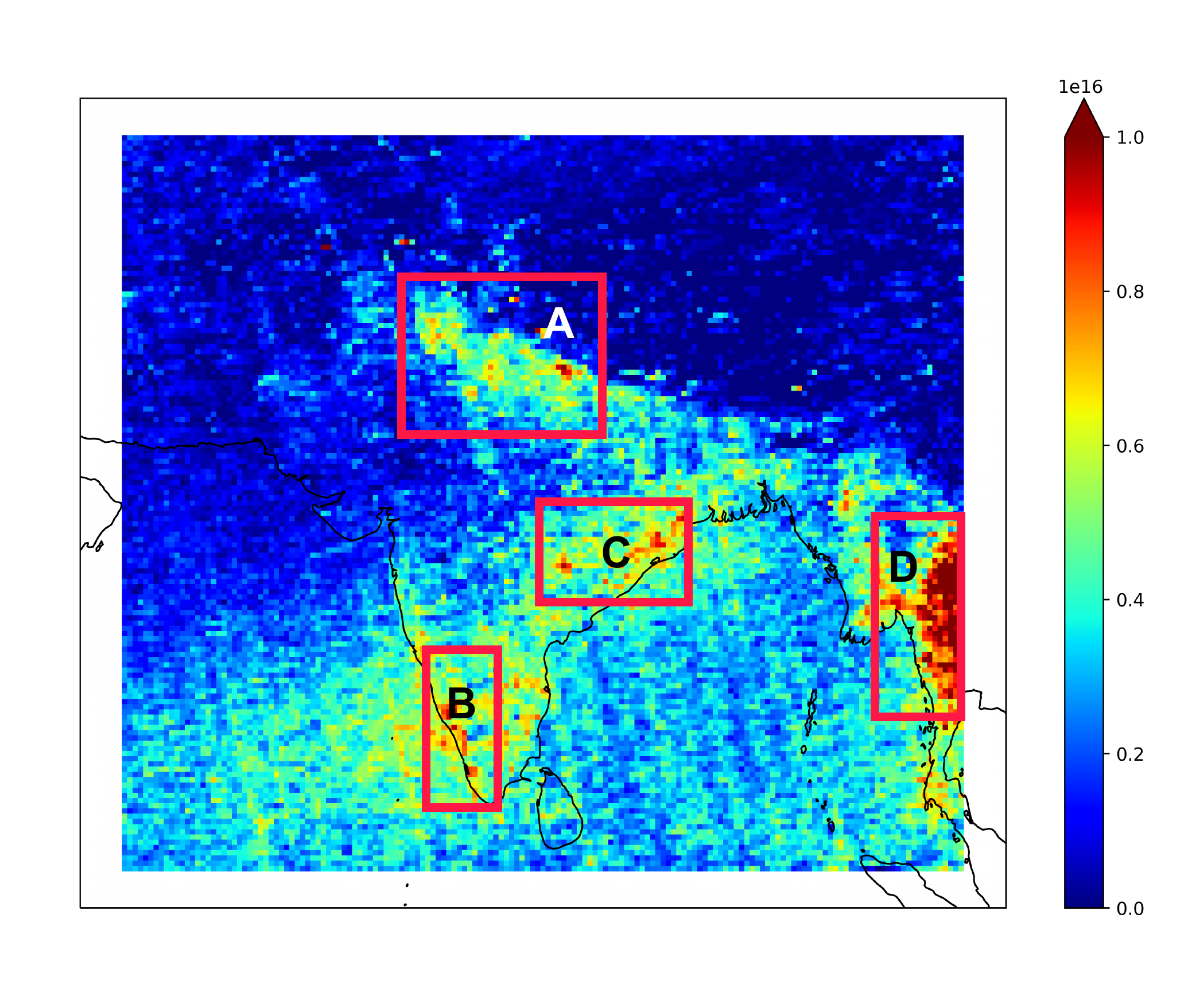}

   \caption{Spatial distribution of average OMI satellite observed formaldehyde. The four boxes show the important hot-spots of HCHO over south Asia. Box 'A' show Indo-Gangetic plain (IGP) region. Box 'B' and 'C' show Western Ghats mountain range and forest region of Indian state Odisha and Chhaattishgarh. Box 'D' shows forest regions of Myanmar. The biogenic and anthropogeic emissions from these regions make them hot-spots for formaldehyde.}
   \label{fig:three}
\end{figure*}

\subsection{Existing gap area and our Contributions }

As we have mentioned earlier that dynamic chemistry models struggle to simulate atmospheric HCHO in terms of both absolute value and geographical variability. Various different chemistry mechanisms with different levels of complexity can be deployed in dynamic chemistry models and the simulated results often differ \cite{zhang2012impact, gupta2015validation, balzarini2015wrf}. Researchers often need to choose suitable chemical mechanism based on available computation resources and objective of the research problem. The learning curve of these models are steep and model results often vary based on the initial conditions used for the model. Here we present a novel and to the best of our knowledge for the first time a data-driven approach to simulate atmospheric HCHO with the use of much less input parameters compared to dynamic models.

\section{Dataset}
Dynamic chemistry models (e.g. WRF-Chem, GEOS-Chem etc.) simulate emission, transportation and chemical reactions along with meteorology. As initial conditions these models need various meteorological parameters (e.g. horizontal and vertical wind, geopotential height, temperature and pressure at different levels etc.) along with concentration and emission information for different chemical species. In this present work we have removed the complexity of dynamic models by only focusing on few input parameters. We we have mentioned earlier that plats emit VOCs at higher temperature as a natural response to heat stress \cite{sharkey2008isoprene}. The oxidation of VOCs to form HCHO is also favoured at higher temperature. Hence we selected atmospheric temperature as one of the input for our model. Photooxidation of VOCs are facilitated by solar radiation, when incoming solar radiation was used and another meteorological input. Atmospheric lifetime of HCHO and \(C_5H_4\), the two most important chemical species in our work, ranges from few minutes to hours. During this short lifetime, long-range transport of these species are unlikely and the concentrations of them will be higher near the emission site \cite{biswas2019simultaneous} and effect of chemical transportation will be less. Hence we have not selected horizontal or vertical winds as input. Biogenic emissions being major contributors for VOCs emission, leaf area index (LAI) with both high and low vegetation were selected as proxies for biogenic emission. Methane is considered to control the background concentration of HCHO and \(C_5H_4\) is the most important biogenic VOC for atmospheric HCHO formation. Hence \(CH_4\) and \(C_5H_4\) concentrations were use as chemistry input for the model. As all higher VOCs contribute to some extend to the atmospheric HCHO formation, we have use total concentration of higher VOCs (except \(CH_4\), \(C_5H_4\) and HCHO) as another input for the model.

We have selected OMI HCHO observations as the target variable. OMI satellite has equitorial overpass time of ~13:30 hrs. Hence OMI satellite observation captures the HCHO signal during midday hours. The HCHO concentration during midday will be dependent on the concentration of its precursors. As we have already discussed that the lifetime of the precursor gases and the HCHO ranges from few minutes to few hours, the HCHO concentration will be dependent on the precursor concentration from few hours prior also. Hence we have used the input data averaged for the period 10:30 hours local time (05:00 hours UTC) to 14:30 hours local time (09:00 hours UTC) daily. We have assumed that the all the input variable will have effect on HCHO concentration for few hours prior to the observation time.

Incoming solar radiation, atmospheric temperature at 2 m height above ground and leaf area index (high and low vegetation) with one hour resolution and 0.1 x 0.1 degree spatial resolution for 05:00-09:00 hours UTC time from ERA5 reanalysis data were used as input (Muñoz Sabater, J., (2021): ERA5-Land hourly data from 1950 to 1980. Copernicus Climate Change Service (C3S) Climate Data Store (CDS). (Accessed on < DD-MMM-YYYY >), 10.24381/cds.e2161bac). Total column \(CH_4\), \(C_5H_8\) and other VOCs (expect \(CH_4\), \(C_5H_8\) and HCHO) from CAMS reanalysis \cite{inness2019cams} with 0.75 x 0.75 degree spatial resolution and three hourly temporal resolution during 03:00, 06:00 and 09:00 hours UTC was used for the study. One daily averaged dataset for all the variable were created from hourly files. OMI satellite HCHO product with 0.1 x 0.1 degree spatial resolution were downloaded from NASA-EARTHDATA portal (Kelly Chance (2019), OMI/Aura Formaldehyde (HCHO) Total Column Daily L3 Weighted Mean Global 0.1deg Lat/Lon Grid V003, Greenbelt, MD, USA, Goddard Earth Sciences Data and Information Services Center (GES DISC), Accessed: [Data Access Date], 10.5067/Aura/OMI/DATA3010). All the dataset were regridded to create dataset with uniform spatial resolution.

\section{Methodology}

The formaldehyde (HCHO) data from OMI satellite is first quality controlled. Firstly, at each grid point, the negative formaldehyde column amount is discarded. Secondly, the formaldehyde column amounts less than the uncertainty at each grid point are cleaned. And, thirdly, the formaldehyde column amounts with data quality flag values of 0 are only retained. The data quality flag is provided by NASA.

For the prediction of formaldehyde (HCHO) from various meteorological, chemical and vegetation species we aggregate the ERA5 reanalysis, CAMS reanalysis and HCHO data from OMI satellite to daily temporal resolution. Then, we compute the total amount of VOCs from CAMS reanalysis using the following equation:

\begin{center}
$VOC = CH_{3}COCH_{3} + C_{2}H_{6} +C_{2}H_{5}OH + \newline C_{2}H_{4} + CH_{3}OH + C_{3}H_{8} + Higher VOCs  $      
\end{center}

The total amount of VOC is used as a precursor while training. We extract all the fields from multiple sources over the South Asian region corresponding to 5-40N, 60-100E. All the datasets are regridded to 0.25 degree spatial resolution corresponding to ERA5 reanalysis. We use different combinations of precursors to target formaldehyde. The combinations are as provided in Table \ref{tab:example} wherein the meteorology corresponds to surface incoming solar radiation and surface air temperature. The data from 2005 to 2014 is used for training, 2015 for validation and 2016 to 2019 for testing. Maximum and minimum values of all the precursors and target formaldehyde are computed from the training dataset and min-max normalization applied to training, validation and testing. 

We use a 3-layer super resolution convolutional neural network (SRCNN) to map from the inputs to target. The kernel size used for the three layers is 9 x 9, 1 x 1 and 5 x 5. We use Adam optimizer with a learning rate of 0.001 and loss function as the mean squared error. The size of the input and target for selected South Asian region comes out as 141 x 161. In order to prevent overfitting, we use early stopping and save the best model following validation loss while training. Training one epoch with a batch size of 64 and total training sample size of 3648 takes 2 seconds on an NVIDIA A100 GPU. The best model is used to perform the test predictions using the precursors from test dataset which are then de normalized to generate the test predictions. The test predictions are compared to the satellite based formaldehyde data to generate the performance statistics.

\begin{figure*}[t]
  \centering
   \includegraphics[width=0.8\linewidth]{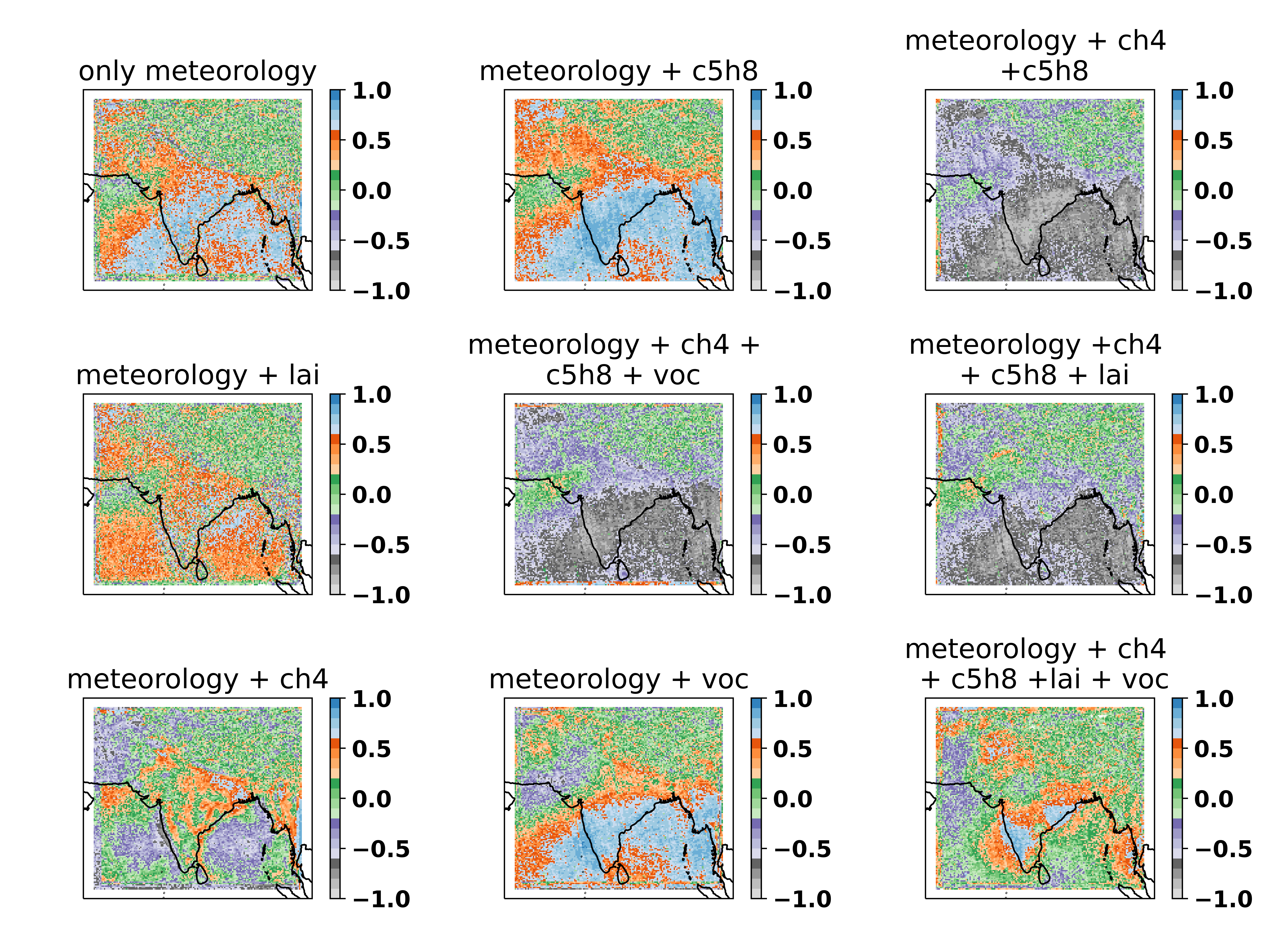}

   \caption{Spatial distribution of Pearson correlation coefficient between modelled and satellite measured formaldehyde.}
   \label{fig:four}
\end{figure*}

\begin{figure*}[t]
  \centering
   \includegraphics[width=0.8\linewidth]{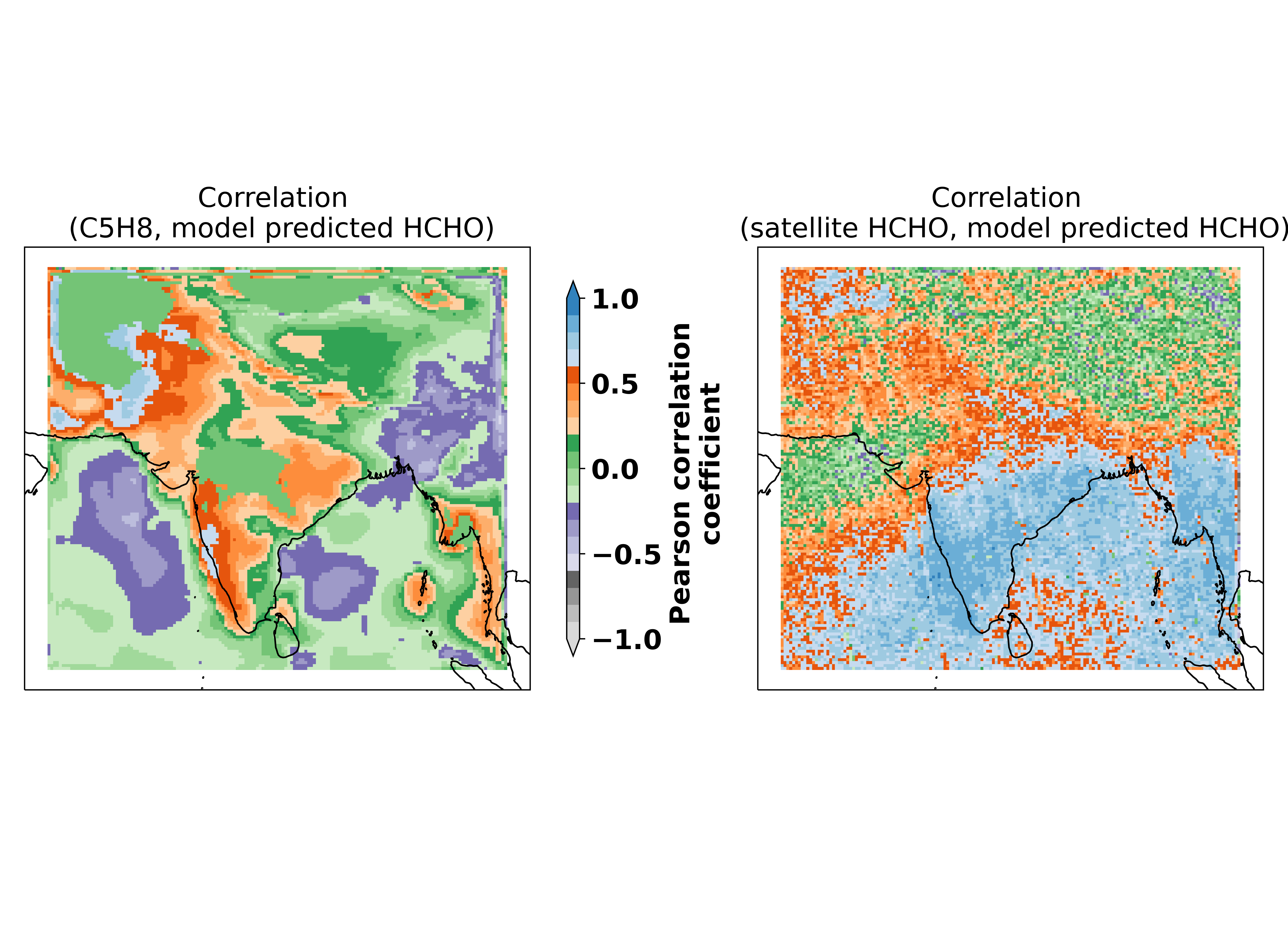}

   \caption{Spatial distribution of correlation between modelled formaldehyde and isoprene (left side) and satellite measured formaldefyde (right side). From the plots it is evident that the modelled formaldehyde is not biased by input isoprene.}
   \label{fig:five}
\end{figure*}



\section{Results}
Figure 3 represents the spatial distribution of average HCHO VCDs over south Asia from OMI-Aura dataset. The plot shows HCHO hotspots over specific regions and the spatial variation quite non-homogeneous. Box 'A' represents the Indo-Gangetic plain region (IGP) from India. IGP region has the highest population density \cite{biswas2019simultaneous} in India with several big cities including Indian capital city Delhi. Anthropogenic emissions from heavy industries, automobiles contribute towards the elevated HCHO levels. In the state of Punjab and Haryana from western IGP region, during post-monsoon (October and November) season open crop residue contributes heavily towards VOC emission \cite{jain2014emission}. Box 'B' shows HCHO hotspot over the Western Ghats mountain ranges from Indian state Kerala and neighbouring places. Western Ghats mountain ranges covered in dense forests which is a large source of biogenic emission. Box 'C' shows another forested region from Indian state Odisha and Chhattisgarh and hence HCHO hotspot. Finally the box 'D' shows the forested region from Myanmar.

Figure 4 shows the spatial correlation plot of OMI HCHO and modelled HCHO from different simulations. Spatial correlations were calculated from monthly average satellite and modelled HCHO. The subplot '1' shows results from the model with only meteorology as input variable. Although the R values (Perason correlation coefficient) over box 'B' and 'C' (refer to Figure 3) are above 0.5, correlation drops over 'A' and 'D' region. Table 1 shows the aggregated Pearson correlation coefficient for the entire model time period. The correlation coefficient for model with only meteorology is equal to 0.3. Subplot '2' shows the spatial correlation form model with meteorology and leaf area index. We can see that the correlation over box 'B' and 'C' drops compared to subplot '1' (i.e. with only meteorology) leading to aggregated Pearson correlation coefficient of 0.19 (Table 1). Subplot '3' shows correlation from model with meteorology and methane. From Table 1 we can see that aggregated correlation coefficient from subplot '3' is negative. Likewise the other simulations with methane as input, in subplot '7' and subplot '8' are also negative. The next low correlation coefficient is for subplot '9' with a value of 0.14. All the simulations with methane as input variable have given poor correlations with satellite HCHO. This is quite surprising as methane is supposed to determine the HCHO background concentrations. Probably the longer lifetime of methane (~ one year) compared to other reactive VOCs (few hours) makes it a poor choice for input variable, as we are not providing any other reaction kinetics information to the model. 

The highest correlation comes from the simulation with meteorology and isoprene with correlation coefficient of 0.47. Subplot 4 shows the spatial correlation plot of HCHO from simulation with meteorology and isoprene. Box 'B', 'C' and 'D' all shows correlation coefficient above 0.5. Even box 'A' shows higher correlation compared to subplot 1. Subplot 6 show spatial correlation from the model with meteorology and VOC as input variable and the aggregated correlation coefficient 0.31. This reinstate the fact that isoprene is one of the most important precursor for atmospheric HCHO. However, in future work we need to run more experiments to understand individual effects of different higher VOCs. 

We wanted to be sure whether the better correlation of HCHO from meteorology and isoprene input is biased by the isoprene input or not. To test this hypothesis we studied correlation of isoprene versus modelled HCHO. Figure 5 shows the correlation of modelled HCHO with isoprene (subplot 1) and OMI HCHO (subplot 2). We can see that the correlation with isoprene is is low and inconsistent compared to the satellite HCHO. This proves that the modelled HCHO is not biased by isoprene distributin and our deep learning algorithm is able to capture the nonlinear relationship between input and target variable.

\section{Conclusions}
We have presented a deep learning based approach to model atmospheric formaldehyde without using complex chemical reaction mechanisms and minimum input variables. We develop a trustworthy model of estimating formaldehyde from the inputs directly involved in the formation or decomposition of atmospheric HCHO. Atmospheric temperature at 2 m height above surface and incoming solar radiation were used as meteorological input variables. Leaf area index was used as proxy for biogenic emission. Methane, isoprene and higher VOCs were considered as chemical precursor of HCHO and used as input variable. Our results demonstrate superior skills relative to the dynamical atmospheric chemistry modelling.

We also estimate the causal linkages and find that using only meteorology as input resulted in correlation coefficient of 0.3 between modelled and satellite HCHO. Although methane is very important for HCHO formation, using methane as input resulted in poor results. Probably higher atmospheric lifetime (hence low reaction) compared to other chemical input makes it an unsuitable choice as an input. Simulation with meteorology and isoprene gives most promising results with overall correlation coefficient of 0.47. 

In future we plan to use advanced deep learning algorithms for HCHO modelling and further improve the performance. We also plan to study the effect of different input variable on model output and finally finding the optimal model with causal explanation. Our aim is to develop a low resource intensive, optimised alternate to dynamical model which can be used by relevant research groups without extensive domain knowledge of atmospheric chemistry.

\begin{table*}
  \centering
  \begin{tabular}{@{}lc@{}}
    \toprule
    Variable combination & Pearson correlation coefficient over South Asia \\
    \midrule
    only meteorology & 0.30 \\
    meteorology + lai & 0.19 \\
    meteorology + ch4 & -0.03\\
meteorology + c5h8 & 0.47 \\
    meteorology + ch4 + c5h8 + voc & -0.39 \\
    meteorology + voc & 0.31\\
meteorology + ch4 +c5h8 & -0.47 \\
    meteorology + ch4 + c5h8 + lai & -0.33 \\
    meteorology + ch4 + c5h8 +lai + voc & 0.14\\
    
    \bottomrule
  \end{tabular}
  \caption{Table for total correlation coefficient between different modelled and satellite formaldehyde. The simulation with meteorology and isoprene as input variable gives the best correlation.}
  \label{tab:example}
\end{table*}

\bibliographystyle{unsrt}  
\bibliography{references}

\end{document}